\documentclass{svjour3}
\smartqed
\usepackage{natbib}
\usepackage{color}
\usepackage{graphicx}
\journalname{Biophysical Reviews}

\renewcommand{\baselinestretch}{2}

\begin{document}

\title{Entropic stabilization of the folded states of RNA due to macromolecular crowding}


\dedication{We dedicate this article to Allen P. Minton, who has been a pioneer in recognizing the 
importance of crowding effects in biology and established methods to quantify them through theory and experiments.}

\author{Natalia A. Denesyuk \and D. Thirumalai}

\institute{Natalia A. Denesyuk \at Biophysics Program, Institute for Physical Science and Technology, University of 
Maryland, College Park, Maryland 20742 \and D. Thirumalai \at Biophysics Program, Institute for Physical Science and 
Technology and Department of Chemistry and Biochemistry, University of Maryland, College Park, Maryland 20742 
               \\ Tel.: +1-301-405-4803\\Fax: +1-301-314-9404\\\email{thirum@umd.edu}}

\date{Received: date / Accepted: date}

\maketitle

\begin{abstract}
We review the effects of macromolecular crowding on the folding of RNA by considering the simplest scenario when 
excluded volume interactions between crowding particles and RNA dominate. Using human telomerase enzyme as an example, 
we discuss how crowding can alter the equilibrium between pseudoknot and hairpin  states of the same RNA molecule --- a 
key aspect of crowder-RNA interactions. We summarize data showing that the crowding effect is significant only if the size 
of the spherical crowding particle is smaller than the radius of gyration of the RNA in the absence of crowding particles.
The implication for function of the wild type and mutants of human telomerase is outlined by using a relationship 
between enzyme activity and its conformational equilibrium. In addition, we discuss the interplay between macromolecular 
crowding and ionic strength of the RNA buffer. Finally, we briefly review recent experiments which illustrate the 
connection between excluded volume due to macromolecular crowding and the thermodynamics of RNA folding.  
\end{abstract}

\clearpage

\section{Introduction}

The cytosol is crowded and replete with macromolecules such as ribosomes, lipids, proteins, and RNA. Estimates show that 
the volume fraction ($\phi$) of these macromolecules, collectively referred to as crowding agents, can exceed 0.2. 
Spontaneous folding of nascent proteins and RNA in the crowded environment can be different from {\it in vitro} experiments
that are typically conducted under infinite dilution. Minton, who has made pioneering contributions in elucidating the 
importance of crowding in biophysics, was the first to recognize that the thermodynamics of folding, association, and 
biochemical reactions can be altered by crowding 
agents~\citep{Minton81Biopolymers,Minton01JBC,Minton05BJ,Hall03BBA,Zhou08ARB}. More recently there has been much interest 
in the study of crowding effects on folding and function of 
proteins~\citep{Cheung05PNAS,Homouz08PNAS,Dhar10PNAS,Elcock10COSB}, which was inspired by the insightful studies initiated 
by Minton. 

Describing even the simplest process, namely, the transition between folded and unfolded states of proteins and RNA under 
crowded conditions is complicated because the nature of the effective interactions between the crowding agents and the
polypeptide or polynucleotide chains is not fully understood. However, to a first approximation, the dominant effect of 
crowding agents is to exclude the molecule of interest from the volume occupied by crowders. If excluded volume interactions 
dominate (an assumption that has to be tested before the theory can be applied to analyze experiments), then the stability 
of the folded state of the protein or RNA is enhanced, compared to the case when $\phi =0$. In this case, the loss 
in entropy of the folded state due to crowding is much less than that of the unfolded state, resulting in the 
stabilization of the folded state. The entropic stabilization of the folded state~\citep{Cheung05PNAS,Minton05BJ} in the 
presence of crowding agents has been firmly established theoretically in a number of studies. 

Most of the studies on the effects of crowding on self-assembly process have been on the folding of proteins. Recently, 
it has been recognized that crowding agents could have a significant impact on the folding of 
RNA~\citep{Pincus08JACS,Kilburn10JACS,Denesyuk11JACS}.  Simple theoretical arguments and coarse-grained simulations were 
used to show that crowding can modestly stabilize RNA secondary structures~\citep{Pincus08JACS}. However, RNA requires 
counterions (Mg$^{2+}$, for example) for tertiary folding. Thus, the effect of macromolecular crowding on tertiary 
structures of RNA may be complicated. Using small angle X-ray scattering measurements it has been shown that, in the 
presence of polyethylene glycol (PEG), the 195 nucleotide {\it Azoarcus} ribozyme is more compact relative 
to $\phi = 0$~\citep{Kilburn10JACS}. It was concluded that excluded volume effects play a dominant role in the 
compaction of RNA in low molecular weight PEG. Interestingly, the transition to the folded state occurs at a lower 
Mg$^{2+}$ concentration in the presence of PEG~\citep{Kilburn10JACS}. Even if excluded volume interactions largely 
determine the stability of the folded states of RNA, a number of variables besides $\phi$, such as size and shape 
of crowding agents, also contribute to the stability of RNA in the presence of inert crowding agents. Thus, 
a systematic study of the influence of macromolecular crowding on RNA is required.

It is known that, in contrast to proteins~\citep{Guo92JCP}, the stability gap separating the native state and low energy 
excitations in RNA is small~\citep{Thirumalai05Biochem}, which implies that external factors (crowding or force) can 
modulate the stabilities and functions of RNA. For example, riboswitches undergo a transition between two distinct 
conformations that have a profound influence on their functions~\citep{Montage08ARB}. In this review article, 
we outline essential aspects of crowder-RNA interactions, with the particular emphasis on how crowding in the cellular 
environment may alter conformational equilibria related to enzymatic function. As a biologically relevant example, we 
discuss crowding effects on the transition between the hairpin (HP) and pseudoknot (PK) conformations (Figure \ref{SS}) in the 
pseudoknot domain of human telomerase RNA, hTR \citep{Theimer03PNAS}. The pseudoknot domain is conserved in different 
organisms and its activity is closely linked to chromosome stability \citep{Blasco03COGD,Chen04PNAS}. 
However, the precise role of the PK and HP conformations of the pseudoknot domain in the context of telomerase activity is 
not known. Mutations that either increase or decrease the stability of the PK conformation result in a reduction in 
telomerase activity \citep{Comolli02PNAS,Theimer05MolCell}. Therefore, it is important to compare the impact of physical 
factors, such as macromolecular crowding, with that of naturally occurring chemical mutations. Our discussion of 
crowding presented in this review is original in two ways. First, we consider the effects of crowding on the conformational 
equilibrium between two folded states. Traditionally, discussions of crowding effects on biomolecules, mostly proteins, 
concern with the stability of the folded (active) state with respect to the unfolded (inactive) state. In addition, we  
establish a quantitative and novel connection between the magnitude of the crowding effect and activity of wild-type and 
mutant enzymes. 

\section {Computational models of RNA and crowders}

Entropic stabilization, the mechanism by which macromolecular crowding stabilizes folded states of proteins and RNA in the 
excluded volume limit, is similar to spatial confinement, in a sense that crowders confine the biomolecule of interest to 
the interstitial space. Although analytical results exist for polymers confined to cavities with simple geometries, it is 
difficult to obtain accurate estimates for nontrivial confinement geometry associated with macromolecular crowding. 
Furthermore, unfolded RNA is a highly nonideal polymer whose conformational ensemble will be determined, among other things, 
by stacking propensities between adjacent bases and by the ionic strength of the buffer. Estimating how the conformational 
space of such a polymer will be reduced even in a simple geometrical confinement is a challenging task. 

On the other hand, simulations have proven to be a useful tool in assessing the magnitude of entropic stabilization in 
response to varying external conditions. Coarse-grained models have been particularly effective in the studies of RNA 
folding, since they do not suffer from computational complexity associated with all-atom force 
fields \citep{Lin08JACS,Whitford09BJ,Cho09PNAS,Feng11JACS,Denesyuk11JACS}. As a specific example of this class of techniques, 
we will be discussing results obtained with a widely appreciated coarse-grained representation of RNA, where each 
nucleotide is modeled by three interactions sites (TIS) --- a phosphate, a sugar and a 
base \citep{Hyeon05PNAS,Cho09PNAS,Denesyuk11JACS}. We developed a force field in conjunction with the TIS model,
which originally included stacking and hydrogen bond interactions as essential components in the stability of RNA 
structures and was thermodynamically accurate in the limit of high ionic strength \citep{Denesyuk11JACS}. The quantitative 
agreement of thermodynamic predictions of this force field with experiments is illustrated in Figure \ref{rC}a. 
Subsequently, we added an electrostatic component to the TIS model to describe the RNA thermodynamics at different 
ionic concentrations (Denesyuk and Thirumalai, to appear elsewhere). The description of electrostatic effects in the model 
is based on Manning's concept of counterion condensation, which posits that counterions condense onto the RNA molecule and 
reduce the charge of phosphate groups from $-e$ to $-Qe$, where $Q<1$ and $e$ is the proton charge. The uncondensed mobile 
ions are treated using the Debye-H{\"u}ckel theory. Therefore, the electrostatic energy of RNA in simulation is computed 
using
\begin{equation}
U_{\rm EL}=\frac{Q^2e^2}{2\varepsilon}\sum_{i,j}\frac{\exp\left(-|{\bf r}_i-{\bf r}_j|/\lambda\right)}{|{\bf r}_i-{\bf r}_j|},
\label{GDH}
\end{equation}
where  $|{\bf r}_i-{\bf r}_j|$ is the distance between two phosphates $i$ and $j$, $\varepsilon$ is the dielectric 
constant of water and $\lambda$ is the Debye-H{\"u}ckel screening length. We showed that, if the reduced phosphate 
charge $Q$ was taken to be
\begin{equation}
Q =\frac{b}{l_{\rm B}},
\label{Manning}
\end{equation}
where $l_{\rm B}$ is the Bjerrum length and $b=4.4$ \r{A}, the measured thermodynamics of a variety of RNA sequences could 
be reproduced well over a wide range of temperatures and monovalent salt concentrations (Denesyuk and Thirumalai, to be 
published).

A generalized Lennard-Jones potential, introduced by \cite{Denesyuk11JACS}, has been successfully employed to model 
interactions of RNA with the spherical crowders of arbitrary size,
\begin{eqnarray}
U_{\rm LJ}(r)&=&\varepsilon\frac{2R_i}{D_0}\left[\left(\frac {D_0}{r + D_0 - D}\right)^{12} 
- 2\left(\frac {D_0}{r + D_0 - D}\right)^{6} + 1\right],\ r\le D, \nonumber \\
U_{\rm LJ}(r)&=&0,\ r > D, 
\label{POT1}
\end{eqnarray}
where $r$ is the distance between the centers of mass of two interacting particles, $D_0$ is the effective penetration 
depth of the interaction, $R_i$ is the radius of an RNA coarse-grained bead, $r_{\rm C}$ is the radius of a crowder,
and $D=R_i+r_{\rm C}$. The ratio $2R_i/D_0$ in Eq.~(\ref{POT1}) is used to scale the interaction strength 
$\varepsilon$ in proportion to the surface contact area. This potential correctly accounts for nonspecific 
surface interactions between spherical crowders representing large macromolecules and individual segments of the 
coarse-grained RNA.

Assessing the magnitude of crowding effect from simulations requires an accurate technique for calculating the
folding free energies. To this end, a thermodynamic method has been proposed (Denesyuk and Thirumalai, to appear 
elsewhere), which does not require a structural order parameter to define folded and unfolded ensembles. To summarize, we 
perform a series of simulations at different temperatures in the range from $T_1$ to $T_2$, where $T_1$ and $T_2$ are on 
the order of the lowest and highest temperatures used in thermodynamic simulations or measurements. Using statistical 
mechanics techniques, we compute from these simulations the free energy of the molecule, $G$, as a function of temperature,
$T$ (Figure \ref{dGillust}). If the population of the unfolded state is negligible at $T_1$, the free energy of the folded 
state at $T_1$, $G_{\rm f}(T_1)$, equals the computed free energy $G(T_1)$. Similarly, $G_{\rm u}(T_2)=G(T_2)$, if the 
population of the folded ensemble is statistically insignificant at the highest temperature $T_2$. Assuming similar 
equalities for the enthalpies and heat capacities of the folded and unfolded state, we can use thermodynamic relationships 
between the free energies, enthalpies and heat capacities to extrapolate $G_{\rm f}(T_1)$ and $G_{\rm u}(T_2)$ to 
intermediate temperatures, at which both folded and unfolded states are populated (asymptotic lines in Figure 
\ref{dGillust}). For any $T$ between $T_1$ and $T_2$, geometric definition of the stability of the folded state with 
respect to the unfolded state, $\Delta G$, is given in Figure \ref{dGillust}.

\section{Crowding effect is negligible for large crowders}

\begin{figure}
\begin{center}
\includegraphics[width=12.0cm,clip]{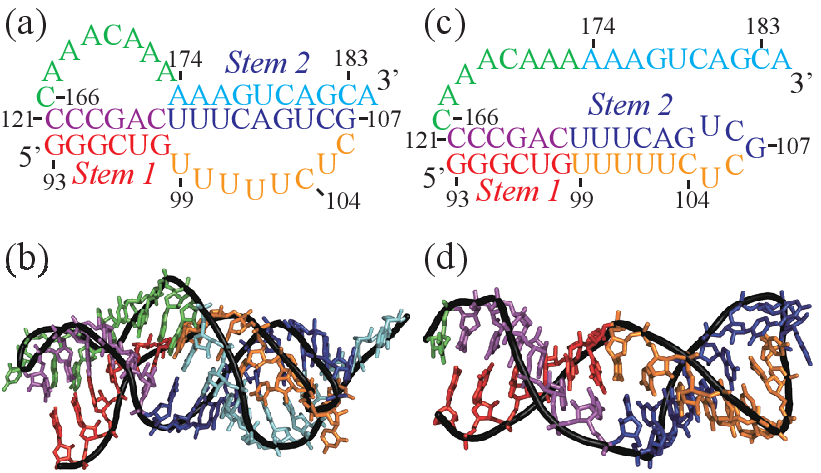}
\end{center}
\caption{Graphic adopted from \cite{Denesyuk11JACS}. Secondary and tertiary structures of the pseudoknot (PK) and hairpin 
(HP) conformations for $\Delta$U177. The sequence of $\Delta$U177 was chosen to match experiments,
in which nucleotides 122--165 and 177 were deleted to facilitate structural studies \citep{Theimer05MolCell}. 
(a) PK secondary structure. (b) PK tertiary structure. (c) HP secondary 
structure. (d) HP tertiary structure. NMR structure of the HP includes residues G93 to C166 only (PDB code 1NA2). To 
quantify the effect of crowders on the PK-HP equilibrium, we added an unstructured tail A167--A184 to the NMR structure 
(panel c). This inclusion ensures that identical RNA sequences are used in simulations of the PK and HP 
conformations.}
\label{SS}
\end{figure}

\begin{figure}
\begin{center}
\includegraphics[width=8.0cm,clip]{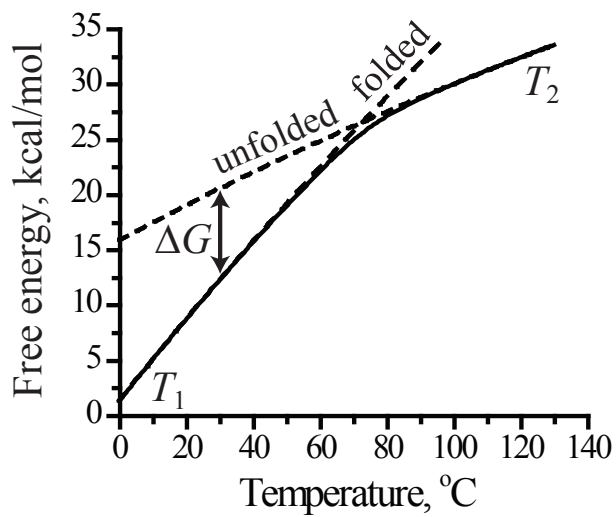}
\end{center}
\renewcommand{\baselinestretch}{2}
\caption{Geometrical definition of the stability $\Delta G$ of the folded state. The solid curve illustrates the total 
free energy of a system as a function of temperature, computed in simulation. The free energies of individual folded 
and unfolded ensembles (dashed curves) are obtained as described in the text.}
\label{dGillust}
\end{figure}

The high density of macromolecules in the cell (volume fractions $\phi \approx 0.2-0.4$) reduces the space available for 
conformational fluctuations. Therefore, macromolecular crowding should result in a shift in the thermodynamic equilibrium 
between the HP and PK states of the hTR pseudoknot domain towards the more compact PK. To assess the extent to which
PK is favored at $\phi \ne  0$, we discuss the simulation results~\citep{Denesyuk11JACS} for the HP and PK states of the 
modified pseudoknot domain, $\Delta$U177 (Figure \ref{SS}). The molecular construct $\Delta$U177 has been examined experimentally 
{\it in vitro} at $\phi$ = 0~\citep{Theimer05MolCell}, and the atomistic structures of its HP and PK conformations are 
available from the Protein Data Bank, codes 1NA2 and 2K96, respectively. Under native conditions, the RNA sequence in 
Figure \ref{SS} will predominantly populate the PK conformation. To obtain adequate statistics on both conformations, two 
independent sets of simulations were carried out, each modeling a limited subset of hydrogen bonds~\citep{Denesyuk11JACS}. 
In the first set of simulations, the RNA sequence could form only those hydrogen bonds that are found in the NMR structure 
of the PK. Similarly, in another set of simulations, only hydrogen bonds from the NMR structure of the HP were included. In
the HP simulations strand C166--A184 remained unbound in the folded state, because no hydrogen bonds could form between 
this strand and the remainder of the molecule. In this way, the interconversion between the PK and HP was eliminated and 
the HP structure could be sampled exhaustively. The stabilities of the PK and HP structures, 
$\Delta G_{\rm PK}=G_{\rm PK, f} - G_{\rm PK, u}$ and $\Delta G_{\rm HP}=G_{\rm HP, f} - G_{\rm HP, u}$, were obtained
from the simulations using the technique illustrated in Figure \ref{dGillust}. Since the unfolded state is effectively defined as a 
high-temperature state (in which all hydrogen bonds are broken) and the RNA sequence is the same in the PK and 
HP simulations, it was assumed that $G_{\rm PK, u} = G_{\rm HP, u}$. Therefore, the stability of the PK 
with respect to HP, $G_{\rm PK, f}-G_{\rm HP, f}$, was computed as $\Delta G_{\rm PK}-\Delta G_{\rm HP}$, without the 
need for explicit simulations of the PK-HP interconversion.

In order to illustrate the essential aspects of crowding effects on RNA, we consider spherical particles with radius 
$r_{\rm C}$. For monodisperse particles, the volume fraction is $\phi = 4 \pi r_{\rm C}^3 \rho/3$, where $\rho$ is 
the number density. Thus, $\phi$ can be changed by increasing or decreasing $\rho$ or by altering the size of the crowding 
particles. In this review, we fix $\phi = 0.3$ and examine the consequences of changing $r_{\rm C}$. Based on general theoretical 
considerations~\citep{Asakura58JPS,Shaw91PRA} it can be shown that, in the colloid limit $r_{\rm C}> R_{\rm G}^0$, the 
crowding agents would have negligible effects on RNA stability. Here, $R_{\rm G}^0$ is the size of RNA in the absence of 
the crowding agent. It is only in the opposite polymer limit, $r_{\rm C}< R_{\rm G}^0$, that the crowding particles 
would affect RNA stability. We therefore expect that the magnitude of the crowding effect should depend on the ratio 
$r_{\rm C}/R_{\rm G}^0$.

\begin{figure}
\begin{center}
\includegraphics[width=12.0cm,clip]{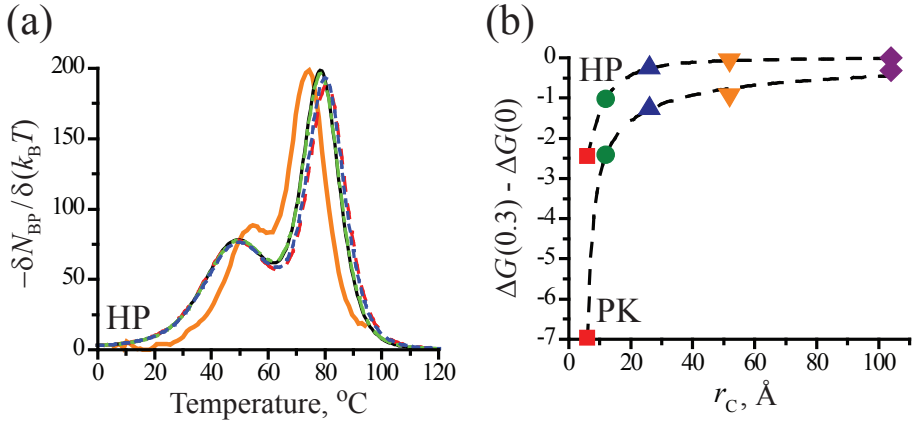}
\end{center}
\caption{(a) Melting profiles of the HP in various crowding environments at 1 M or higher monovalent salt concentration. 
Black solid curve: without crowders. Red dashed curve (the model {\it E. coli} mixture): $\phi=0.11$ of crowders
with $r_{\rm C}=104$ \r{A}, $\phi=0.11$ of crowders with $r_{\rm C}=52$ \r{A} and $\phi=0.08$ of crowders with 
$r_{\rm C}=26$ \r{A}. Green dashed-dotted curve: $\phi=0.08$ of crowders with $r_{\rm C}=26$ \r{A}. Blue dotted curve: 
$\phi=0.3$ of crowders with $r_{\rm C}=26$ \r{A}.
The thick orange curve is the experimental UV data at 
200 mM KCl from Figure 2b in \cite{Theimer03PNAS}, divided by $5.53\times10^{-5}$. The two peaks indicate melting of stems 
1 and 2 of the HP (Figure \ref{SS}c). The peak positions (melting temperatures) and the overall width of the melting curve (melting range) serve as a measure of agreement 
between theory and spectroscopic data. (b) Changes in stability (kcal/mol) of the HP and PK at 37 $^{\circ}$C due to 
crowders at $\phi=0.3$, as a function of the crowder radius $r_{\rm C}$.}
\label{rC}
\end{figure}

In Figure \ref{rC}a we show the HP melting profile, taken to be the negative derivate of the number of intact base pairs 
$N_{\rm BP}$ with respect to $k_{\rm B} T$, for the crowder radius $r_{\rm C}=26$ \r{A}~\citep{Denesyuk11JACS}. Such 
crowders are larger than the radius of gyration of strand G93--C121 in the unfolded state, $R_{\rm G}^0=20$ \r{A} 
(Figure \ref{SS}c). As discussed above, large crowders have minimal effects on the melting of the HP even at $\phi=0.3$. For a 
fixed $\phi$, the average distance between two spherical crowders will increase with the crowder size. If the unfolded 
hairpin can easily fit in the interstitial space, the folding/unfolding transition will not be affected significantly by 
the presence of crowders. For $\phi=0.3$ and the crowder radius $r_{\rm C}=26$ \r{A}, which is only slightly larger than 
$R_{\rm G}^0$, the increase $\Delta T$ in the melting temperature is 1.5 $^{\circ}$C for stem 1 of the HP and is 
negligible for stem 2 (Figure \ref{rC}a). Further increase in $r_{\rm C}$ results in $\Delta T\approx0$ for both stems (data not 
shown).

Figure \ref{rC}a also shows the melting profile of the HP in a ternary mixture of crowders, containing volume fractions 
$\phi=0.11$, 0.11 and 0.08 of particles with $r_{\rm C}=104$ \r{A}, 52 \r{A} and 26 \r{A}, 
respectively~\citep{Denesyuk11JACS}. The sizes and volume fractions of individual components in the model mixture correspond 
to the ribosome, large enzymatic complexes and relatively small individual proteins, found in {\it E. coli}. Because all 
the values of $r_{\rm C}$ in the {\it E. coli} mixture are larger than $R_{\rm G}^0$, we expect only small changes in the 
melting profile of the HP (Figure \ref{rC}a). For the total volume fraction of 0.3, the melting temperature of the HP stem 1 
increases only by 2 $^{\circ}$C with respect to $\phi=0$ (Figure \ref{rC}a). Interestingly, the effect of the {\it E. coli} 
mixture is similar in magnitude to that of a monodisperse suspension with $r_{\rm C}=26$ \r{A} and $\phi=0.3$. In contrast,
a monodisperse suspension with $r_{\rm C}=26$ \r{A} and $\phi=0.08$, which is equivalent to the smallest particle component 
in the mixture, has negligible effect on the melting of the HP (Figure \ref{rC}a). 

To summarize, the crowding effect of polydisperse mixtures is largely the effect of the smallest particle component, 
but taken at the total volume fraction of the mixture. As we discuss in the next section, the excess stability of the 
folded state due to crowding decreases nonlinearly with the radius of the crowding particle $r_{\rm C}$. We therefore 
propose that, for crowding in the cellular environment, the main role of large macromolecules will be to increase the 
effective volume fraction of the relatively small macromolecules. 

\section{Role of crowder size in the PK-HP equilibrium}

The sensitivity of the crowding effect to the relative sizes of RNA and crowders is at the basis of the equilibrium
shift in the hTR pseudoknot domain. Figure \ref{rC}b shows the change in stability of the HP and PK at 37 $^{\circ}$C 
induced by monodisperse crowders for different crowder radii $r_{\rm C}$ ($\phi=0.3$). As anticipated by arguments given 
above, the magnitude of the excess stability $\Delta G(0.3)-\Delta G(0)$ is small if $r_{\rm C}/R_{\rm G}^0>1$ and 
increases sharply for $r_{\rm C}/R_{\rm G}^0<1$. Note that the crowding effect is larger for the PK for all values of 
$r_{\rm C}$ (Figure \ref{rC}b), indicating an equilibrium shift towards this conformation. The discussed change in the 
PK-HP relative stability is of entropic origin. The unfolded ensembles of the PK and HP are equivalent, as discussed above,
and will therefore be depleted by macromolecular crowding to a similar degree. The compact folded PK is, to a first 
approximation, unaffected by crowding. However, the folded HP contains an unbound strand C166--A184 (Figure \ref{SS}), 
whose loose conformations will be restricted in a crowded environment. Therefore, the excess stability of the PK with 
respect to HP is due to a partial suppression of the HP folded ensemble by macromolecular crowding.

The crowder radius $r_{\rm C}=12$ \r{A} corresponds to the size of an average protein {\it in vivo}. 
For $\phi=0.3$ and $r_{\rm C}=12$ \r{A}, we have $\Delta G_{\rm PK}(0.3)-\Delta G_{\rm PK}(0)=-2.4$ kcal/mol and 
$\Delta G_{\rm HP}(0.3)-\Delta G_{\rm HP}(0)=-1.0$ kcal/mol, which amounts to the relative stabilization of the PK 
conformation by $-1.4$ kcal/mol (Figure \ref{rC}b). Below we analyze this value in the context of standard changes in the 
PK stability caused by mutations.

\section{Implications for function}

\begin{figure}
\begin{center}
\includegraphics[width=10.0cm,clip]{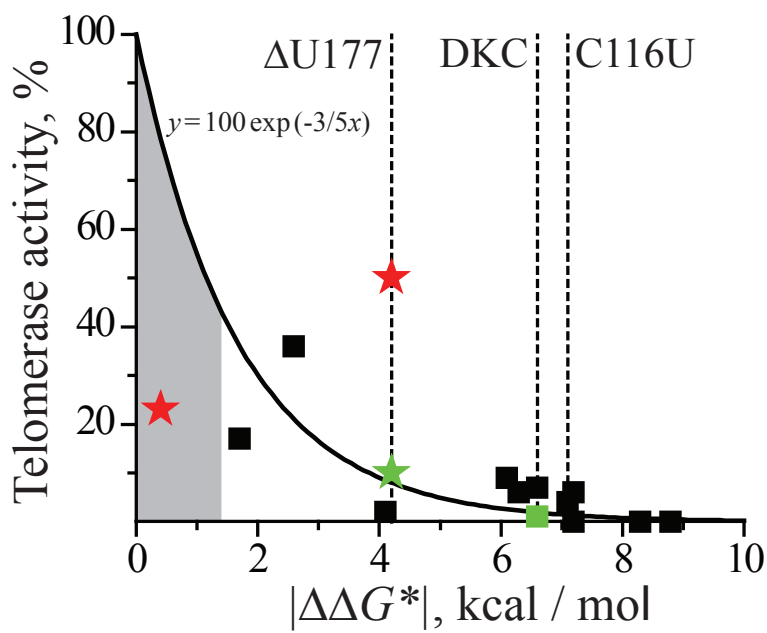}
\end{center}
\caption{Graphic adopted from \cite{Denesyuk11JACS}. Activity of mutant telomerase normalized to wild-type activity (100\%), as a 
function of the magnitude of the stability difference between mutant and wild-type pseudoknots. Three mutants, 
$\Delta$U177, DKC and C116U, are explicitly marked. The {\it in vitro} data for destabilizing (black squares) and 
stabilizing (red stars) mutations are from Table 2 in \cite{Theimer05MolCell}. Green symbols show the 
{\it in vivo} data for $\Delta$U177 and DKC from \cite{Comolli02PNAS}. The experimental data for destabilizing mutations is fit to 
the exponential function (solid curve). Grey area marks the range of stability differences that can be accommodated by 
crowding.}
\label{Activity}
\end{figure}

As mentioned in the Introduction,  changes in the relative stability of the HP and PK conformations of the hTR pseudoknot 
domain compromise the enzyme activity. The estimate of the crowding effect in a typical cellular environment,
$\Delta\Delta G=-1.4$ kcal/mol, allows us to assess the extent to which macromolecules could regulate telomerase 
activity. Experimental data on hTR mutants~\citep{Comolli02PNAS,Theimer05MolCell} clearly demonstrate that enzyme 
activity decreases when plotted as a function of $|\Delta\Delta G^*|=|\Delta G^*_{\rm PK}(0)-\Delta G_{\rm PK}(0)|$, 
where $\Delta G^*_{\rm PK}(0)$ and $\Delta G_{\rm PK}(0)$ are the stabilities of mutant and wild-type pseudoknots at 
$\phi=0$ (Figure \ref{Activity}). The majority of mutations destabilize the PK, $\Delta\Delta G^*>0$ (black squares in Figure \ref{Activity}) and only two 
mutants have $\Delta\Delta G^*<0$ (red stars in Figure \ref{Activity}). For destabilizing mutants the reduction in activity, 
$\alpha$, was shown by \cite{Denesyuk11JACS} to follow the exponential dependence, $\alpha=\exp (-0.6\Delta\Delta G^*)$ 
(thick curve in Figure \ref{Activity}). The naturally occurring destabilizing mutations DKC and C116U have been linked to diseases 
dyskeratosis congenita and aplastic anemia, respectively~\citep{Vulliamy02Lancet,Fogarty03Lancet}. The DKC and the 
stabilizing $\Delta$U177 mutations have been studied {\it in vivo} (green symbols in Figure \ref{Activity}), as well as 
{\it in vitro}. In both cases, mutant telomerase {\it in vivo} was found to be significantly less active than the 
corresponding construct {\it in vitro}, suggesting that a number of factors determine the activity of telomerase 
{\it in vivo}. 

Although macromolecular crowding enhances the stability of the PK state, the crowding effect ($\Delta\Delta G=-1.4$ kcal/mol) is less than 
the stability changes caused by mutations. In Figure \ref{Activity} the grey area marks the domain of potential mutants with 
$\Delta\Delta G^*>0$, whose activity may be completely restored by macromolecular crowding. All experimentally studied 
mutants fall outside the marked domain, including the two disease related mutants DKC and C116U. Nevertheless, due to the 
strong dependence of enzyme activity on $\Delta\Delta G^*$, the effect of crowding on telomerase function may be 
significant. We estimate that the activity of telomerase can be up- or down-regulated by more than two-fold in response 
to density fluctuations in its immediate environment. Furthermore, due to the expected dynamical heterogeneities in cells, 
there will be variations in enzyme activity in different cell regions.

\section{Crowding effects on RNA at different ionic strengths}

\begin{figure}
\begin{center}
\includegraphics[width=12.0cm,clip]{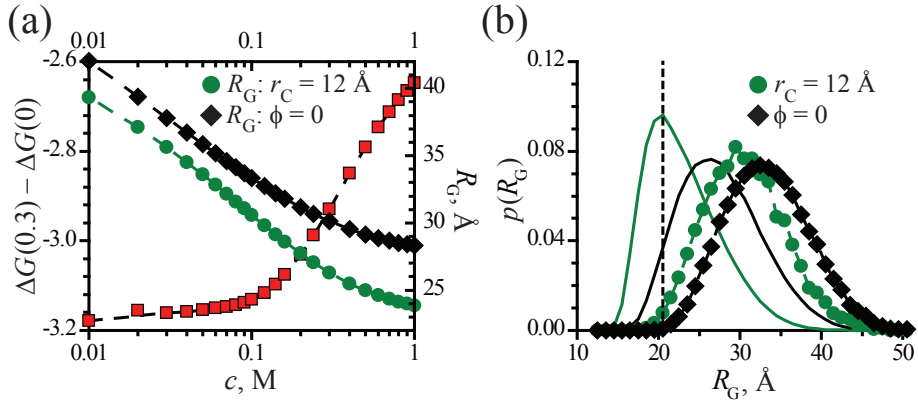}
\end{center}
\caption{(a) The radius of gyration of the unfolded PK (right axis), as a function of the monovalent ion concentration $c$,
for $\phi=0$ (black diamonds) and $\phi=0.3$, $r_{\rm C}=12$ \r{A} (green circles). Red squares show the excess 
stability of the PK due to crowding, $\Delta G_{\rm PK}(0.3)-\Delta G_{\rm PK}(0)$, for $r_{\rm C}=12$ \r{A} (left axis).
(b) Probability distributions $p(R_{\rm G})$ of the radius of gyration of the unfolded PK for $\phi=0$ and $c=1$ M 
(black solid line), for $\phi=0$ and $c=0.1$ M (black diamonds), for $\phi=0.3$, $r_{\rm C}=12$ \r{A} and $c=1$ M 
(green solid line), and for $\phi=0.3$, $r_{\rm C}=12$ \r{A} and $c=0.1$ M (green circles). The vertical dashed line indicates 
the smallest size of RNA conformations that will be perturbed by crowders with $\phi=0.3$ and $r_{\rm C}=12$ \r{A}.}
\label{Salt}
\end{figure}

Entropic stabilization mechanism~\citep{Cheung05PNAS} implies that crowding increases the stability of the folded state by reducing the
population of expanded conformations in the unfolded state. Therefore, we expect that the magnitude of the crowding 
effect will be sensitive to the ionic strength of the RNA buffer, since the latter determines the size of the 
unfolded RNA. The quantitative discussion above assumed the limit of high ionic strength. As the buffer ionic concentration $c$ is lowered, the screening of the negative charge on the RNA 
sugar-phosphate backbone becomes less efficient, which in turn increases the mean radius of gyration of conformations in 
the unfolded state. The function $R_{\rm G}(c)$ for the unfolded PK is shown in Figure \ref{Salt}a in the absence (black diamonds) 
and presence (green circles) of crowding. The same general trend is observed in both cases, with the $R_{\rm G}$ values 
being consistently smaller when crowders are present for the entire range of $c$. In accord with our predictions, the 
crowder-induced stabilization of the folded PK becomes more significant at low ionic strengths (red squares in Figure \ref{Salt}a). 
Interestingly, the stabilization effect increases rapidly upon lowering $c$ from 1 M to 0.1 M, but shows little change
when $c$ is lowered further to below 0.1 M. The underlying reason for such behavior can be traced to the probability 
distributions $p(R_{\rm G})$ in the unfolded state (Figure \ref{Salt}b). For a given crowder solution, we can identify a typical 
size of the cavity which will be free of any 
crowders. If the radius of gyration of RNA conformations is such that they fit into the cavity, these conformations will 
not be perturbed by crowding. On the other hand, the population of conformations with $R_{\rm G}$ larger than the typical 
cavity size will be significantly depleted by crowding. For $\phi=0.3$ and $r_{\rm C}=12$ \r{A}, we can infer the size of 
a standard empty cavity from the distributions $p(R_{\rm G})$ at high ionic strength ($c=1$ M, Figure \ref{Salt}b). Note 
that $p(R_{\rm G})$ decreases for $R_{\rm G}>20$ \r{A} when crowders are present (green solid line in Figure \ref{Salt}b), 
but $p(R_{\rm G})$ increases with $R_{\rm G}$ around 20 \r{A} in the absence of crowders (black solid line in 
Figure \ref{Salt}b). This indicates that the crowders significantly perturb the unfolded conformations with $R_{\rm G}$ 
larger than 20 \r{A}, which can serve as an upper estimate of the smallest RNA size affected by crowders. When $c$ 
decreases, the distribution $p(R_{\rm G})$ shifts to larger $R_{\rm G}$, increasing the fraction of the unfolded 
conformations affected by crowders. At $c=0.1$ M, all statistically significant values of $R_{\rm G}$ in the unfolded 
state fall within the range $R_{\rm G}>20$ \r{A}, both for $\phi = 0$ (black diamonds in Figure \ref{Salt}b) and 
for $\phi=0.3$ (green circles in Figure \ref{Salt}b), so that the entire distribution $p(R_{\rm G})$ is depleted due to 
crowding. This explains why the crowding-induced stabilization is almost constant below 0.1 M, even if the average 
$R_{\rm G}$ continues to increase rapidly all the way to 0.01 M (symbols in Figure \ref{Salt}a).

\section{RNA becomes compact as $\phi$ increases}

\begin{figure}
\begin{center}
\includegraphics[width=10.0cm,clip]{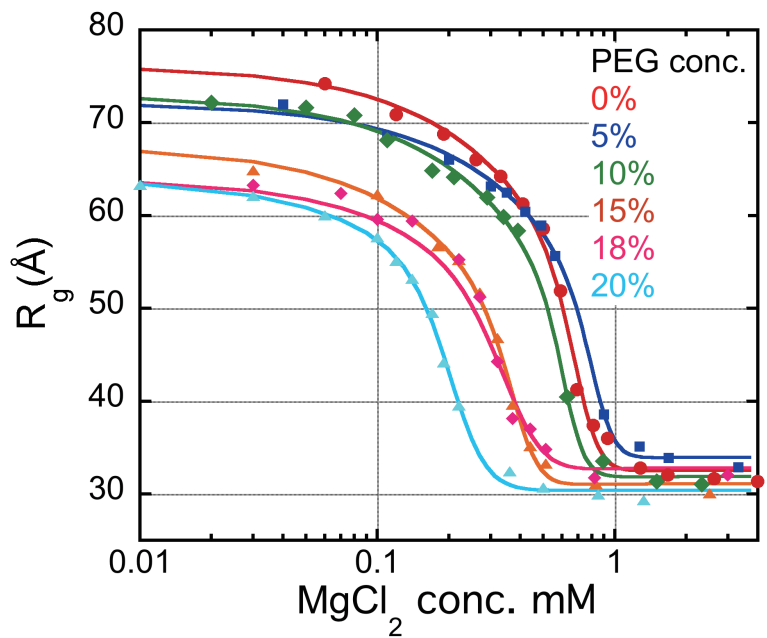}
\end{center}
\caption{Graphic adopted from \cite{Kilburn10JACS}. Small angle X-ray scattering measurements of the radius of gyration of the 
{\it Azoarcus} ribozyme for different concentrations of Mg$^{2+}$ ions and PEG.}
\label{Azoarcus}
\end{figure}

The reduction in conformational space accessible to RNA should increase with $\phi$, for a fixed $r_{\rm C}$. Thus, we 
expect that $R_{\rm G}$ should decrease as $\phi$ increases. This is precisely what is observed in 
experiments~\citep{Kilburn10JACS}, which show that at all concentrations of Mg$^{2+}$ the {\it Azoarcus} ribozyme becomes 
more compact as the volume fraction of the crowding agent (PEG) increases (Figure \ref{Azoarcus}). Interestingly, the midpoint of the folding 
transition $c_{\rm m}$ --- the concentration of Mg$^{2+}$ at which the folded and unfolded states of the {\it Azoarcus} 
ribozyme have equal populations --- also decreases as $\phi$ increases (Figure \ref{Azoarcus}). This finding can be readily 
explained in terms of the entropic stabilization mechanism~\citep{Cheung05PNAS} and suggests that, to a first approximation,
PEG behaves as an inert hard sphere crowding agent. Based on our considerations from the previous section, we predict that 
the shift in $c_{\rm m}$ due to crowding will also depend on the concentration of monovalent counterions in the RNA buffer,
a prediction that is amenable to experimental test. In addition, it would be of interest to perform experiments at a fixed 
$\phi$ but varying $r_{\rm C}$, which can be changed by decreasing or increasing the molecular weight of PEG. 

\section{Conclusions}

The phenomena discussed here illustrate just one aspect of crowding effects on RNA. There are other physical and chemical
factors that could determine how crowding modulates RNA stability, and hence its function. The interplay between electrostatic interactions, 
hydration of phosphate groups and crowding particles, as well as the shape of crowding particles, could potentially present
a much more nuanced picture than the simplest case considered here. Even though these effects are important and warrant 
further study, the analysis presented here is important in establishing the maximum increase in stability that can be 
realized when excluded volume interactions dominate. From this perspective, our conclusion that the crowding effects 
cannot fully restore telomerase activity {\it in vivo} is robust. In future work, it will be important to connect the 
consequences of crowding effects on RNA folding with the RNA activity under cellular conditions. 

\begin{acknowledgements} 
We are grateful to Sarah Woodson for providing Figure 6. Our work 
was supported by a grant from the National Science Foundation (CHE 09-10433).
\end{acknowledgements}
Conflict of Interest: None

\clearpage

\clearpage

\clearpage

\clearpage

\clearpage

\clearpage

\end{document}